%% file: stochHYPE-QAPL.tex
\documentclass[copyright,creativecommons]{eptcs}
\providecommand{\event}{QAPL 2011} 
\usepackage{breakurl}              

\providecommand{\urlalt}[2]{\href{#1}{#2}}
\providecommand{\doi}[1]{doi:\urlalt{http://dx.doi.org/#1}{#1}} 

\usepackage{amssymb}
\usepackage{amsmath}
\usepackage{pepa}
\usepackage{graphicx}
\usepackage{rotating}
\usepackage{stmaryrd}
\usepackage{url}
\usepackage[mathscr]{eucal}

\newtheorem{definition}{Definition}
\newtheorem{remark}{Remark}
\newtheorem{example}{Example}

\input{defQAPL}

\renewcommand{\TS}[1]{\mathcal{T\hspace*{-4.0pt}S}_{#1}}
\renewcommand{\TC}[1]{\mathcal{T\hspace*{-1.5pt}C}_{#1}}
\renewcommand{\TD}[1]{\mathcal{T\hspace*{-3.0pt}D}_{#1}}

\newcommand{\pc}{{\,:\,}}
\newcommand{\Sys}{\mathit{Sys}}
\newcommand{\TempCtrl}{\mathit{TempCtrl}}
\newcommand{\Heat}{\mathit{Heat}}
\newcommand{\Time}{\mathit{Time}}
\newcommand{\Sun}{\mathit{Sun}}
\newcommand{\Cool}{\mathit{Cool}}
\newcommand{\Shade}{\mathit{Shade}}
\newcommand{\Dwnldr}{\mathit{Dwnldr}}
\newcommand{\dw}{\mathit{dw}}

\newcommand{\W}{\mathcal{W}}

\def\S{\mbox{\large $\rhd\!\!\!\!\!\lhd$}}

\title{HYPE with stochastic events}
\author{Luca Bortolussi
\institute{Department of Maths and Computer Science,
University of Trieste.} 
\email{luca@dmi.units.it} 
\and
Vashti Galpin \qquad\qquad Jane Hillston
\institute{Laboratory for Foundations of Computer Science,
University of Edinburgh}
\email{Vashti.Galpin@ed.ac.uk, Jane.Hillston@ed.ac.uk} \and
}
\def\titlerunning{HYPE with stochastic events}
\def\authorrunning{L. Bortolussi, V. Galpin and J. Hillston}

\begin{document}
\maketitle

\begin{abstract}
The process algebra HYPE was recently proposed as a fine-grained
modelling approach for capturing the behaviour of hybrid systems. In
the original proposal, each flow or influence affecting a variable
is modelled separately and the overall behaviour of the system then
emerges as the composition of these flows.  The discrete behaviour
of the system is captured by instantaneous actions which might be
urgent, taking effect as soon as some activation condition is
satisfied, or non-urgent meaning that they can tolerate some
(unknown) delay before happening. In this paper we refine the notion
of non-urgent actions, to make such actions governed by a
probability distribution.  As a consequence of this we now give HYPE
a semantics in terms of Transition-Driven Stochastic Hybrid
Automata, which are a subset of a general class of stochastic
processes termed Piecewise Deterministic Markov Processes.
\end{abstract}

\input{intro}

\input{hype}

\input{TDSHA}

\input{mapping}

\input{related}

\input{conc}

\input{biblio}
\end{document}

%% file: defQAPL.tex
\usepackage{tikz}
\usetikzlibrary{automata}

\newcommand{\ev}[1]{\text{\underbar{#1}}}
\newcommand{\sev}[1]{\overline{\text{#1}}}
\newcommand{\ssev}[1]{\textrm{#1}}

\newcommand{\Ev}{\mathcal{E}}

\newcommand{\ec}{\mathop{\mathit{ec}}}

\newcommand{\EC}{\mathit{EC}}

\newcommand{\iname}[1]{\mathit{#1}}

\newcommand{\const}{\mathit{const}}
\newcommand{\linear}{\mathop{\mathit{linear}}}

\renewcommand{\Ac}{\mathcal{A}}

\newcommand{\iv}{\mathop{\mathit{iv}}}

\newcommand{\ID}{\mathit{ID}}
\newcommand{\IN}{\mathit{IN}}
\newcommand{\IT}{\mathit{IT}}

\newcommand{\assign}[1]{\llbracket #1 \rrbracket}
\newcommand{\eval}[1]{\llbracket #1 \rrbracket}
\newcommand{\V}{\mathcal{V}}

\newcommand{\ConSys}{\mathit{ConSys}}
\newcommand{\Con}{\mathit{Con}}
\newcommand{\CSys}{\mathcal{C}_\mathit{Sys}}

\newcommand{\bbR}{\mathbb{R}}

\newcommand{\calM}{\mathcal{M}}

\newcommand{\is}{\mathsf{is}}
\newcommand{\es}{\mathsf{ev}}

\newcommand{\vr}[1]{\mathbf{#1}}
\newcommand{\TC}[1]{\mathcal{TC}_{#1}}
\newcommand{\TD}[1]{\mathcal{TD}_{#1}}
\newcommand{\TS}[1]{\mathcal{TS}_{#1}}
\newcommand{\TDSHA}[1]{\mathcal{T}_{#1}}
\newcommand{\TDSHAtuple}[1]{(Q_{#1},\vr{X_{#1}},\TC{#1},\TD{#1},\TS{#1},\mathrm{init}_{#1},\mathcal{E}_{#1})}
\newcommand{\initial}[1]{\mathrm{init}_{#1}}
\newcommand{\modes}[1]{Q_{#1}}
\newcommand{\vrX}[1]{\mathbf{X_{#1}}}
\newcommand{\events}[1]{\mathcal{E}_{#1}}
\newcommand{\tprod}[1]{\otimes_{#1}}

\newcommand{\mode}[1]{q_{#1}}
\newcommand{\initmode}[1]{q_{#1}^{init}}
\newcommand{\stoich}[1]{\mathbf{s}_{#1}}
\newcommand{\rf}[1]{f_{#1}}
\newcommand{\exit}[1]{q_{1}^{#1}}
\newcommand{\enter}[1]{q_{2}^{#1}}
\newcommand{\priority}[1]{w_{#1}}
\newcommand{\guard}[1]{g_{#1}}
\newcommand{\res}[1]{r_{#1}}
\newcommand{\e}[1]{e_{#1}}
\newcommand{\inp}[1]{\mathrm{inp}_{#1}}

\newcommand{\act}{\mathit{act}}
\newcommand{\rs}{\mathit{res}}

%% file: intro.tex
\section{Introduction}\label{sec:intro}

Process algebras have been successfully applied to the analysis and
verification of a wide variety of systems over the last thirty
years.  Although initially focused on semantic issues of concurrent
programming, their compositional style and ability to support a
number of different analysis techniques has extended their use into
many application domains.  In the realm of quantified analysis
stochastic process algebras, in which actions are associated with a
randomly distributed delay, have been  used to study the dynamics of
diverse systems ranging from the performance of software systems
\cite{Hill96} to
the biochemical signalling in living cells \cite{CiocCH:09a}.  Such
an analysis is inherently based on a discrete state view of the
system with an underlying semantics which is generally a continuous
time Markov chain (CTMC).  In contrast, recently, process algebras
have been used to study situations of collective dynamics in which
a fluid approximation of the discrete state space is used to arrive
at a semantics in terms of sets of ordinary differential equations
(ODEs) \cite{HillH:05a,TribTGH:10a}.

Hybrid behaviour arises in a variety of systems, both engineered and
natural.  Such systems combine elements of both the approaches
outlined above as the system will undergo periods of continuous
evolution, governed by ODEs, punctuated by discrete events which can
alter the course of subsequent continuous evolution.  Consider a
thermostatically controlled heater. The continuous variable is air
temperature, and the discrete events are the switching on and off of
the heater by the thermostat in response to the air temperature
\cite{TuffTCT:01a}.
Another example would be a genetic regulatory network, such as the
Repressilator \cite{ElowEL:00a,GHB08a}, in which genes can be switched on
or off by interactions with their environment (more precisely, with
transcription factor proteins). The behaviour of such systems can be
regarded as a collection of sets of ODEs, the discrete events
shifting the dynamic behaviour from the control of one set of ODEs
to another.  This is the approach taken with hybrid automata
\cite{HenzH:96a}. Given the previous success of capturing discrete
and continuous scenarios with process algebras in the past it is
therefore natural to consider process algebras for hybrid settings.

A number of process algebras for describing hybrid systems have
appeared in recent years~\cite {KhadK:06a}, substantially differing
in the approaches taken relating to syntax, semantics, discontinuous
behaviour, flow-determinism, theoretical results and availability of
tools. However, they are all similar in their approach in that the
dynamic behaviour of each subcomponent must be fully described with
the ODEs for the subcomponent given explicitly in the syntax of the
process algebra, before the model can be constructed. What
distinguishes HYPE~\cite{CONCUR09} is that it captures behaviour at
a fine-grained level, composing distinct flows or influences which
act on the continuous variables of the system. At a superficial
level this removes the need to explicitly write ODEs in the process
algebra syntax. Instead the dynamic behaviour emerges, via the
semantics of the language, when these  elements are composed.
Moreover the use of flows as the basic elements of model
construction has advantages such as ease and simplification of
modelling. This approach assists the modeller in allowing them to
identify smaller or local descriptions of the model and then to
combine these descriptions to obtain the larger system. The explicit
controller also helps to separate modelling concerns.

In the original definition of HYPE, discrete actions  are termed
\emph{events} and are always considered \emph{instantaneous}
although some are subject to an activation condition which will
determine when that instantaneous jump occurs.  Most events are
conditioned on the values of continuous variables which are evolving
in the system and will be triggered when the activation condition
becomes true; such events are termed \emph{urgent}.  Many systems
also respond to events which are not so tightly tied to the
continuous evolution of the system and may appear to occur randomly.
In the original definition of HYPE such actions were given an
undefined activation condition, denoted $\bot$ and termed
\emph{non-urgent}.  However if we wish to carry out quantified
analysis of the constructed models such events may be regarded as
underspecified since we capture no information about their potential
firing.  Thus here we seek to refine this notion of non-urgent
events, by introducing \emph{stochastic actions}.  These actions
will have an activation condition which is a random variable,
capturing the probability distribution of the time until the event
occurs.  Thus these event still occur non-deterministically and are
not directly linked to the values of continuous variables, but they
are now quantified and so the models admit quantitative analysis.

This small modification substantially enriches the class of
underlying mathematical processes which capture the behaviour of
systems modelled in HYPE\@.  Previously we gave HYPE a semantics in
terms of hybrid automata \cite{HenzH:96a}.  Now we give a semantics
in terms of Piecewise Deterministic Markov Processes (PDMPs)
\cite{STOC:Davis:1993:PDMP}, using the richer class of automata,
Transition Driven Stochastic Hybrid Automata (TDSHA) as an
intermediary. Due to space constraints, in this paper we will only
show how to associate a TDSHA for a given HYPE model. Mapping TDSHAs
to PDMPs can be done along the lines
of~\cite{SB:Bortolussi:2009:HybridsCCPLattice}.

TDSHA have also been used in \cite{BortBGHT:10a} to define a hybrid
semantics for PEPA, a well-known stochastic process algebra \cite{Hill96}.
That application of TDSHA is rather different from the one presented
here. In \cite{BortBGHT:10a}, we construct a hybrid system approximating
the behaviour of the CTMC associated with a PEPA model by the
standard semantics, using just continuous flows and stochastic
events. HYPE, by contrast, is a process algebra expressly designed
to model hybrid system, hence it deals with both instantaneous and
stochastic events.

The rest of this paper is organised as follows. In
Section~\ref{sec:HYPE} we briefly recall the basic notions of HYPE
by means of a running example, explaining how to extend it in the
stochastic setting in Section~\ref{sec:stochHYPE}.
Sections~\ref{sec:TDSHA} and \ref{sec:HYPEtoTDSHA} are devoted to
recall the definition of TDSHA and to describe how to construct a
TDSHA for a given HYPE model. Finally,
Sections~\ref{seC:relatedWork} and \ref{sec:conc} discuss related
work and draw final conclusions.

%% file: hype.tex
\section{HYPE Definition}\label{sec:HYPE}

In this section we  recall the definition  of non-stochastic HYPE by
way of a running example. More details about the language can be
found in~\cite{CONCUR09,HYPE-journal}.

We consider an orbiter which travels around the earth and needs to
regulate its temperature to remain within operational limits.  It
has insulation but needs to use a heater at low temperatures and at
high temperatures it can erect a shade to reflect solar radiation
and reduce temperature. Its HYPE model, is given in
Table~\ref{orbiter}. The whole system is described by
$\TempCtrl$, and it is composed of two pieces: an
uncontrolled system $\Sys$ and a controller $\Con$,
plus some additional information.

HYPE modelling is centered around the notion of \emph{flow}, which is
some sort of influence continuously modifying one variable. Both the
strength and form of a flow can be changed by \emph{events}. In our
example, we identify four flows affecting the temperature, modeled
by the variable $K$. One is due to thermodynamic cooling, one is due
to the heater, one is due to the heating effect of the sun and one
is due to the cooling effect of the shade.

Flows are described by the \emph{uncontrolled system}, a composition
of several sequential subcomponents, each modelling how a specific
flow is changed by events. For instance, in Table~\ref{orbiter}, the
subcomponent $\Heat$ describes the heating system, which
reacts to the events turning it on and off ($\ev{on}$ and
$\ev{off}$). The tuple $(h,r_h,\const)$ following event $\ev{on}$,
is called an \emph{activity} or an \emph{influence} and describes
how the heater affects the temperature when it is working: $h$ is
the name of the influence, which provides a link to the target
variable of the flow ($K$ in our example), $r_h$ is the strength of
the influence and $\const$ is the influence type, identifying the
functional form of the flow (which is specified separately by the
interpretation $\assign{\const} = 1$). When the heater is turned
off, the influence $(h,r_h,\const)$ is replaced by $(h,0,\const)$,
i.e. the influence strength of the heater becomes zero. The other
subcomponents affecting temperature are $\Shade$, $\Sun$,
and $\Cool(K)$, while $\Time$ keeps track of the
flow of time. States of a HYPE model are collections of influences,
one for each influence name, defining a set of ordinary differential
equations describing the continuous evolution of the system. For
instance, $(h,r_h,\const)$ contributes to the ODE of $K$ with the
addend $r_h\assign{\const}=r_h$.


\begin{figure}[!t]

\begin{eqnarray*}
\TempCtrl & \rmdef & \Sys \sync{M}
\ev{init}.\Con \quad \text{with} \quad M = \{ \ev{init}, \ev{on},
\ev{off}, \ev{up}, \ev{down}, \ev{light}, \ev{dark} \}. \\
\\
\Sys & \rmdef & (((\Heat \sync{\{\ev{init}\}} \Shade)
\sync{\{\ev{init}\}} \Sun) \sync{\{\ev{init}\}} \Cool(K))
\sync{\{\ev{init},\ev{light},\ev{dark}\}} \Time \\
\\
\Heat & \rmdef & \ev{on}\pc(h,r_h,\const).\Heat +
              \ev{off}\pc(h,0,\const).\Heat +
              \ev{init}\pc(h,0,\const).\Heat \\
\Shade & \rmdef & \ev{up}\pc(d,-r_d,\const).\Shade
+ \ev{down}\pc(d,0,\const).\Shade + \\
& & \ev{init}\pc(d,0,\const).\Shade \\
\Sun & \rmdef & \ev{light}\pc(s,r_s,\const).\Sun
+ \ev{dark}\pc(s,0,\const).\Sun 
+ \ev{init}\pc(s,0,\const).\Sun \\
\Cool(K) & \rmdef & \ev{init}\pc(c,-1,\linear(K)).\Cool(K)\\
\Time & \rmdef & \ev{light}\pc(t,1,\const).\Time
          \! + \! \ev{dark}\pc(t,1,\const).\Time \! + \!
            \ev{init}\pc(t,1,\const).\Time \\
\\
\Con & \rmdef & \Con_h \sync{\emptyset} Con_d \sync{\emptyset} \Con_s \\
\Con_h & \rmdef & \ev{on}.\ev{off}.\Con_h \quad \Con_d \:\rmdef\:
\ev{up}.\ev{down}.\Con_d \quad \Con_s \:\rmdef\:
\ev{light}.\ev{dark}.\Con_s 
\end{eqnarray*}

$\begin{array}{rclrcl}
\iv(t) & = & T & \iv(h) & = & \iv(d) \: = \: iv(s) \: = \: iv(c) \: = \:
K \\
\\
\ec(\ev{init}) & = & \multicolumn{1}{l}{(true,(K'=t_0\wedge T'=0))} \\
\ec(\ev{off}) & = & (K\geq k_1,true) \quad & \ec(\ev{on}) & = & (K\leq k_2,true) \\
\ec(\ev{up}) & = & (K\geq k_3,true) \quad & \ec(\ev{down}) & = & (K\leq k_4,true) \\
\ \ \ \ 
\ec(\ev{light}) & = & (T=12,true) \quad & \ec(\ev{dark}) & = & (T=24,T'=0) \\
\end{array}$
\caption{Orbiter model in HYPE.}\label{orbiter}
\end{figure}


The controller $\Con$, instead, is used to impose causality on
events, either due to nature (such as the alternation of day and
night) or by design. For instance, $\Con_h$ expresses the fact that
the heater can be turned off only if it is on. Events happen when
certain conditions are met by the system.  These \emph{event
conditions} are specified by a function $ec$, assigning to each
event a \emph{guard} or \emph{activation condition} (stating when a
transition can fire) and a \emph{reset} (specifying how variables
are modified by the event). For example, $\ec(\ev{on}) = (K\leq
k_2,true)$ states that the heater is turned on when the temperature
falls below a threshold $k_2$ and no variable is modified and
$\ec(\ev{dark}) = (T=24,T'=0)$ states that the event $\ev{dark}$
happens after 24 hours and resets the clock $T$ to zero.
Events in HYPE are urgent, meaning that they fire as soon as their
guard becomes true. HYPE has also non-urgent events, whose guard is
denoted by  $\bot$. They can happen after an unconstrained,
non-deterministic time delay.

A full HYPE model is given by
$(\ConSys,\V,\IN,\IT,\Ev,\Ac,\ec,\iv,\EC,\ID)$, where $\ConSys$
is the controlled system, $\V$ is the set of continuous variables,
$\Ev$ is the set of events, $\EC$ is the set of event conditions,
$\ec:\Ev\rightarrow\EC$ associates event conditions to events, $\IN$
is a set of influence names, $\IT$ is a set of influence types,
$\Ac$ is a set of possible influences, $\iv: \IN \rightarrow \V$
maps influence names to variable names, and $\ID$ associates a
real-valued function with each influence type. Formally, the
semantics of HYPE is defined via structured operational semantics
\cite{CONCUR09,HYPE-journal}, which is then interpreted in terms
of hybrid automata.


\subsection{HYPE with stochastic events}\label{sec:stochHYPE}

We now consider how the HYPE language can be  enriched with
stochastic transitions, namely events which are not triggered by
particular values of system variables but according to a random
variable, whose distribution may depend on system variables or may
be independent.  These transitions may be considered as a
generalisation of the non-urgent transitions which were previously
specified with the event condition $\bot$.  In the simplest case
they will correspond to an event which occurs after an exponentially
distributed delay with constant fixed rate.

To illustrate the use of stochastic transitions we  consider an
extension of our previous orbiter example.  We now suppose that as
well as monitoring its own temperature in order to regulate it and
maintain correct operation, the orbiter is also collecting
temperature data. These data are periodically downloaded to earth.
The instigation of the download comes from a control room on earth
and is outside the control of the orbiter. This will be governed by
an exponential distribution with a fixed, constant rate. Between
downloads, data will accumulate deterministically at a constant
rate. When a download is commenced its duration will depend on the
amount of data which has currently accumulated and will thus be an
exponential distribution with a fixed parameter which depends on a
system variable.  It is possible to also imagine a download rate
which is dependent on the current temperature of the orbiter, which
would be an exponential distribution with a variable rate.

We assume that the system variable recording  the amount of data
currently stored on the orbiter is $D$.  The value of $D$ is
governed by two influences representing the accumulation and
downloading of data respectively.  These are $\delta_1=(\dw,r,\const);
\delta_0 = (\dw,0,\const)$ respectively. Clearly both these correspond
to a single influence name $\dw$, with $\iv(\dw) = D$.

The two events that modify the status of the influence $\dw$ are
$\sev{request}$ and $\sev{completed}$, and they are stochastic. We
model this fact by assuming that their activation condition is a
rate function, depending on the value of continuous variables, which
is the parameter of the exponential distribution governing their
firing time. Resets, instead, behave as for instantaneous
transitions. Hence,

$$\ec(\sev{request}) = (\lambda_r,true)\;\;\; \quad
\ec(\sev{completed}) = \left(\frac{\lambda}{\mu + D},D' = 0\right).$$
The form of the rate function for $\sev{completed}$ guarantees that
its rate is $\lambda/\mu$ when $D=0$ and goes monotonically to zero as
$D$ goes to infinity (i.e. expected time of the event is minimal
when there is no data, and grows linearly with $D$).
Here $\lambda/\mu$  represents the maximum downloading speed (which
is achieved when there is no data to collect), while $\mu$ controls the
amount of data required to halve the download speed.

Then, the full orbiter model is obtained by adding one more
component to the uncontrolled system of previous section.
\[\Dwnldr \rmdef \ev{init}\pc\delta_1.\Dwnldr +
\sev{request}\pc\delta_0.\Dwnldr +
\sev{completed}\pc\delta_1.\Dwnldr \]
Furthermore, the downloading events are controlled by the following
controller, synchronizing them with the rest of the system:
\[ \Con_\dw \rmdef \sev{request}.\sev{completed}.\Con_\dw \]
Note how the compositionality of HYPE allows us to extend models in
a simple and natural way.


From a syntactic point of view, a stochastic HYPE model is described
by a tuple
$(\ConSys,\V,\IN,\IT,$ $\Ev_d,\Ev_s,\Ac,\ec,\iv,\EC,\ID)$ in a similar
fashion to HYPE. The main
difference with respect to non-stochastic HYPE is that events are
separated into two disjoint sets, $\Ev_d$ and $\Ev_s$, the
instantaneous and the stochastic events,
respectively\footnote{Events $\ev{a}\in\Ev_d$ are indicated by
underlined letters, while events $\sev{a}\in\Ev_s$ are denoted by
letter with a line above them. A generic event, either stochastic or
instantaneous, is indicated with $\ssev{a}\in\Ev=\Ev_s\cup\Ev_d$.}.
Furthermore, event conditions are different between instantaneous
and stochastic events.  From a semantic  point of view, instead, the
semantics of stochastic HYPE will be defined by associating a
(Transition-Driven) Stochastic Hybrid Automaton to each HYPE model,
as described in the next sections. We now give the formal definition of
a stochastic HYPE model which consists of a controlled system together
with the appropriate sets and functions.

\begin{definition}
A \emph{stochastic HYPE model} is a tuple 
$(\ConSys,\V,\IN,\IT,\Ev_d,\Ev_s,\Ac,\ec,\iv,\EC,\ID)$ where
\begin{itemize}
\item $\ConSys$ is a controlled system as defined below.
\item $\V$ is a finite set of variables.
\item $\IN$ is a set of influence names and $\IT$ is a set of influence
type names.
\item $\Ev_d$ is the set of instantaneous events of the form $\ev{a}$ and $\ev{a}_i$.
\item $\Ev_s$ is the set of stochastic events of the form  $\sev{a}$ and
$\sev{a}_i$.
\item $\Ac$ is a set of activities of the form
$\alpha(\W) = (\iota,r,I(\W)) \in (\IN \times
\mathbb{R} \times \IT)$ where $\W \subseteq \V$.
\item $\ec:\Ev \rightarrow \EC$ maps events to event conditions. Event
conditions are pairs of activation conditions and resets. Resets
are formulae with free variables in $\V \cup \V'$. Activation
conditions for instantaneous events $\Ev_d$ are formulas  with
free variables in $\V$ and the second, while for stochastic
events of $\Ev_s$, they are functions
$f:\bbR^{|\V|}\rightarrow\bbR^+$.
\item $\iv: \IN \rightarrow \V$ maps influence names to
variable names.
\item $\EC$ is a set of event conditions.
\item $\ID$ is a collection of definitions consisting
of a real-valued function for each influence type name
$\assign{I(\W)} = f(\W)$ where the variables in
$\W$ are from $\V$.
\item $\Ev$, $\Ac$, $\IN$ and $\IT$ are pairwise disjoint.
\end{itemize}
\end{definition}

\begin{definition}
A \emph{controlled system} is constructed as follows.
\begin{itemize}
  \item \emph{Subcomponents} are defined by $C_s(\W) = S$, where $C_s$
  is the \emph{subcomponent name} and $S$ satisfies the grammar
  $S'::= \ssev{a}:\alpha.C_s\ |\ S'+S'$
  ($\ssev{a}\in\Ev=\Ev_d\cup\Ev_s$, $\alpha\in\Ac$), with the
  free variables of $S$ in $\W$.
  \item \emph{Components} are defined by $C(\W) = P$, where $C$
  is the \emph{component name} and $P$ satisfies the grammar
  $P'::= C_s(\W)\ |\ C(\W)\ |\ P'\sync{L}P'$, with
  the free variables of $P$ in $\W$ and $L\subseteq\Ev$.
  \item An \emph{uncontrolled system} $\Sigma$ is defined according
  to the grammar $\Sigma' ::= C_s(\W)\ |\ C(\W)\ |\
  \Sigma'\sync{L}\Sigma'$, where $L\subseteq\Ev$ and $\W \subseteq \V$.
\item \emph{Controllers} only have events:
$M ::= \ev{a}.M\ |\ 0 \ |\ M + M$ with $\ev{a} \in \Ev$ and $L
\subseteq \Ev$ and $\Con ::= M\ |\ \Con \smash{\sync{L}} \Con$.
\item A \emph{controlled system} is
$\ConSys ::= \Sigma \smash{\sync{L}} \ev{init}.\Con$ where $L
\subseteq \Ev$. The set of controlled systems is $\CSys$.
\end{itemize}
\end{definition}

\begin{remark}\label{rem:wellDefined}
All HYPE models that will be considered in the paper comply with the
definition of well-defined HYPE models, given
in~\cite{HYPE-journal}. Essentially, each subcomponent must be a
self-looping agent of the form $S = \sum_{i=1}^k
\ssev{a}_i\pc\alpha_i.S + \ev{init}\pc\alpha.S$, with each $\alpha_i$ of
the form $(\iname{i}_S,r_i, I_i)$, where $\iname{i}_S$ is an
influence name appearing only in subcomponent $S$. Furthermore,
synchronization must involve all shared events. In the following, we
will also assume that all events appearing in the uncontrolled
system appear also in the controller. If an event $\ssev{a}$ is not
subject to any control (apart from its guard), then we always add a
controller of the form  $\Con = \ssev{a}.\Con$.
\end{remark}

%% file: TDSHA.tex
\section{Transition-driven Stochastic Hybrid Automata}\label{sec:TDSHA}

We  now present Transition-Driven Stochastic Hybrid Automata,
introduced in~\cite{SB:Bortolussi:2009:HybridsCCPLattice}, a
formalization of stochastic hybrid automata putting emphasis on
transitions, which can be either discrete (corresponding to
instantaneous or stochastic jumps) or continuous (representing flows
acting on system's variables). This formalism can be seen as an
intermediate layer in defining the stochastic hybrid semantics of
HYPE. In fact, TDSHA can be mapped to Piecewise Deterministic Markov
Processes~\cite{STOC:Davis:1993:PDMP}, so that their dynamics can be
formally specified in terms of the latter. Due to space constraints,
we will not provide a formal treatment of this construction, and
refer the reader to~\cite{SB:Bortolussi:2009:HybridsCCPLattice} for
further details.  In this context, we will also consider a different
notion of TDSHA-product, in which transitions can be synchronized on
their labelling events.

\begin{definition}\label{def:TSHS}
A Transition-Driven Stochastic Hybrid Automaton (TDSHA) is a tuple
$\TDSHA{} =
(Q,\vr{X},\TC,\TD,$ $\TS,\mathrm{init},\mathcal{E})$, where
\begin{itemize}
\item $\modes{}$ is a finite set of \emph{control modes}.

\item $\vrX{} = \{X_1,\ldots,X_n\}$ is a set of real valued
\emph{system's variables}\footnote{Notation: the time
  derivative of $X_j$ is denoted by $\dot{X_j}$, while the value
of $X_j$ after a change of mode is indicated by $X_j'$}.

\item $\TC{}$ is the set of \emph{continuous transitions or flows},
whose elements $\tau$ are triples
$(\mode{\tau},\stoich{\tau},\rf{\tau})$, where $\mode{\tau}\in
\modes{}$ is a mode, $\stoich{\tau}$ is a vector of size
$|\vrX{}|$, and $\rf{\tau}:\bbR^n\rightarrow \bbR$ is a
(sufficiently smooth) function.

\item $\TD{}$ is the set of \emph{instantaneous transitions}, whose
elements $\delta$ are tuples of the form
$(\exit{\delta},\enter{\delta},\guard{\delta},\res{\delta},$
$\priority{\delta},\e{\delta})$.  The transition goes from mode
$\exit{\delta}$ to mode $\enter{\delta}$ and it is labeled by
$\e{\delta}\in\events{}$.  $\priority{\delta}\in \bbR^+$ is the
weight of the edge, used to solve non-determinism among two or more
active transitions. The \emph{guard} $\guard{\delta}$ is a first-order
formula with free variables from $\vr{X}$, representing the
\emph{closed set} $G_{\delta} = \{\vr{x}\in\bbR^n~|~\guard{}[\vr{x}]\}$,
while the \emph{reset} $\res{\delta}$  is a conjunction of formulae
of the form $X'=\rho(\vr{X})$, for some variables of the system.
Variables not appearing in $\res{}$ are not modified, so that the
formula $true$ corresponds to the identity reset.

\item $\TS{}$ is the set of \emph{stochastic transitions}, whose
elements $\eta$ are tuples of the form $\eta =
(\exit{\eta},\enter{\eta},\guard{\eta},\res{\eta},$ $\rf{\eta},\e{\eta})$,
where $\exit{\eta}$, $\enter{\eta}$, $\guard{\eta}$, $\e{\eta}$,
and $\res{\eta}$ are as for transitions in $\TD{}$, while
$\rf{\eta}:\bbR^n\rightarrow \bbR^+$ is the rate function giving
the instantaneous probability of taking transition $\eta$. We require
transitions labeled by the same event to have \emph{consistent}
rates: if $\e{\eta_1} = \e{\eta_2}$, then $\rf{\eta_1}=\rf{\eta_2}$.

\item $\events{}$ is a finite set of event names, labelling discrete
transitions.  $\events{}$ can be partitioned into
$\events{d}\cup\events{s}$, such that all events labelling instantaneous
transitions belong to $\events{d}$, while all events labelling
stochastic transitions are from $\events{s}$.

\item $\initial{}$ is a pair $(\initmode{},\inp{})$, with
$\initmode{}\in\modes{}$ and $\inp{}$ a quantifier-free first
order formula with free variables in $\vrX{}$, representing a
point in $\bbR^n$. $\initial{}$ describes the initial state of
the system.
\end{itemize}
\end{definition}

\paragraph{Dynamics of TDSHA.}
In order to formally define the dynamical evolution of TDSHA, we can
map them into a well-studied model of Stochastic Hybrid Automata,
namely Piecewise Deterministic Markov
Processes~\cite{STOC:Davis:1993:PDMP}.
We just sketch now some ideas about the dynamical behaviour of
TDSHA.
\begin{itemize}
  \item Within each discrete mode $\mode{}\in \modes{}$, the system follows the
  solution of a set of ODE, constructed combining the effects of
  the continuous transitions $\tau$ acting on mode $q$. The
  function $\rf{\tau}(\vr{X})$ is multiplied by the vector
  $\stoich{\tau}$ to determine its effect on each variable and
  then all such functions are added together, so that the ODEs
  in mode $q$ are $\dot{\vr{X}} = \sum_{\tau~|~\mode{\tau} = q}
  \stoich{\tau}\cdot \rf{\tau}(\vr{X})$.
  \item Two kinds of discrete jumps are possible. Stochastic
  transitions are fired according to their rate, similarly to
  standard Markovian Jump Processes. Instantaneous transitions,
  instead, are fired as soon as their guard becomes true. In
  both cases, the state of the system is reset according to the
  specified reset policy.\footnote{Note that the formula
$\res{}$ defines a function from $\bbR^n$ into $\bbR^n$, which
will be also denoted throughout by $\res{}$.} Choice among
several active stochastic or instantaneous transitions is
performed probabilistically proportionally to their rate or
priority.
  \item A trace of the system is therefore a sequence of instantaneous and random jumps
  interleaved by periods of continuous evolution.
\end{itemize}

\paragraph{Product of TDSHA.}
We define now a notion of product of TDSHA which, differently from
the one introduced in~\cite{SB:Bortolussi:2009:HybridsCCPLattice},
allows also the synchronization of discrete transitions on specific
events. In order to do this, we must take care of resets, requiring
that synchronized transitions do not reset the same variable in
different ways. Hence, we say that two transitions
$\delta_1,\delta_2$ (either both discrete or both stochastic) are
\emph{reset-compatible} if and only if
$\e{\delta_1}\neq\e{\delta_2}$ or
$\res{\delta_1}\wedge\res{\delta_2}\neq \mathit{false}$. Two TDSHA
are reset-compatible if and only if all their discrete or stochastic
transitions are pairwise reset-compatible. A similar notion is
required for the initial conditions: Two TDSHA are
\emph{init-compatible} if and only if, given initial conditions
$\initial{1} = (\initmode{1},\inp{1})$ and $\initial{2} =
(\initmode{2},\inp{2})$, then $\inp{1}\wedge\inp{2}\neq
\mathit{false}$.

\begin{definition}\label{def:TDSHAproduct}
Let $\TDSHA{i}=\TDSHAtuple{i}$, $i=1,2$ 
two reset-compatible and init-compatible TDSHA, and let $S\subseteq
\events{1}\cap\events{2}$ be the synchronization set. The
$S$-product $\TDSHA{}  = \TDSHA{1}\tprod{S} \TDSHA{2}=
(Q,\vr{X},\TC,\TD,$ $\TS,\mathrm{init},\mathcal{E})$
is defined by
\begin{enumerate}
  \item $\modes{} = \modes{1}\times\modes{2}$;
  \item $\vrX{} = \vrX{1}\cup\vrX{2}$;
  \item $\events{} = \events{1}\cup\events{2}$;
  \item $\initial{} = (\initmode{},\inp{})$, where $\initmode{} =
  (\initmode{1},\initmode{2})$ and $\inp{} =
  \inp{1}\wedge\inp{2}$.
  \item The set of continuous transitions in a mode
  $\mode{}=(\mode{1},\mode{2})$ contains all continuous
  transitions of $\mode{1}$ and all those of $\mode{2}$:
  $$\TC{}
  = \left\{\left((\mode{1},\mode{2}),\stoich{},\rf{} \right)~|~\mode{1}\in\modes{1},\mode{2}\in\modes{2},(\mode{1},\stoich{},\rf{})\in\TC{1}\vee (\mode{2},\stoich{},\rf{})\in\TC{2} \right\}$$
  \item The set of instantaneous transitions $\TD{}$ is the union of
  non-synchronized instantaneous transitions $\TD{NS}$ and of
  synchronized ones $\TD{S}$, where

{\renewcommand{\arraystretch}{0.5}
  $\begin{array}{rcl}
  \quad \TD{NS} & = &
  \Big\{\left((\mode{1},\mode{2}),(\mode{1}',\mode{2}'),\guard{},\res{},\priority{},\e{}\right)~|~\\
   & & \quad (\mode{i},\mode{i}',\guard{},\res{},\priority{},\e{})\in\TD{i}\wedge\mode{j}=\mode{j}'\in\modes{j}\wedge i\neq j \wedge\e{}\not\in
  S\Big\},
  \end{array}$

   and

$\begin{array}{rcl}
  \quad \TD{S} &=& \Big\{\left((\mode{1},\mode{2}),(\mode{1}',\mode{2}'),\guard{1}\wedge\guard{2},\res{1}\wedge\res{2},\min\{\priority{1},\priority{2}\},\e{}\right)~|~ \\
   & & \quad
  (\mode{1},\mode{1}',\guard{1},\res{1},\priority{1},\e{})\!\in\!\TD{1}\wedge
  (\mode{2},\mode{2}',\guard{2},\res{2},\priority{2},\e{})\!\in\!\TD{2}
   \wedge\e{}\!\in\!S\Big\}.
\end{array}$
}

During synchronization, we apply a conservative policy by taking
the conjunction of guards and resets, and by taking the minimum
of weights.

  \item The set of stochastic transitions is defined similarly as $\TS{} = \TS{NS}\cup\TS{S}$, with

{\renewcommand{\arraystretch}{0.5}
  $\begin{array}{rcl}
\quad \TS{NS} & = &
  \Big\{\left((\mode{1},\mode{2}),(\mode{1}',\mode{2}'),\guard{},\res{},\rf{},\e{}\right)~|~\\
   & &\quad (\mode{i},\mode{i}',\guard{},\res{},\rf{},\e{})\in\TS{i}\wedge\mode{j}=\mode{j}'\in\modes{j}\wedge i\neq j\wedge\e{}\not\in
  S\Big\},
  \end{array}$

and

$\begin{array}{rcl}
\quad \TS{S} & = &
  \Big\{\left((\mode{1},\mode{2}),(\mode{1}',\mode{2}'),\guard{1}\wedge\guard{2},\res{1}\wedge\res{2},\rf{},\e{}\right)~|~\\
  & & \quad (\mode{1},\mode{1}',\guard{1},\res{1},\rf{},\e{})\in\TS{1}\wedge
  (\mode{2},\mode{2}',\guard{2},\res{2},\rf{},\e{})\in\TS{2}\wedge\e{}\in
  S\Big\}.
  \end{array}$
}

  In the synchronization of stochastic transitions, we use the
fact that the rate is the same for all transitions labeled by
the same event, as required by the consistency condition.
\end{enumerate}
\end{definition}

%% file: mapping.tex
\section{Mapping HYPE to TDSHA}\label{sec:HYPEtoTDSHA}

The mapping from HYPE to TDSHA works compositionally, by associating
a TDSHA with each single subcomponent and with each piece of the
controller, then taking their synchronized product according to the
synchronization sets of the HYPE system. Guards, rates, and resets
of discrete edges will be incorporated in the TDSHA of the
controller, while continuous transitions will be extracted from the
uncontrolled system.

Consider a HYPE model
$(\ConSys,\V,\IN,\IT,\Ev_d,\Ev_s,\Ac,\ec,\iv,\EC\!,\ID)$ with
$\ConSys ::= \Sigma \smash{\sync{L}} \ev{init}.\Con$. Here $\Sigma$
is the uncontrolled system and $\Con$ is the controller. In the
following, we will refer to the activation condition and the reset
of an event $\ssev{a}\in\Ev$ by $\act(\ssev{a})$ and
$\rs(\ssev{a})$, respectively.


\paragraph{TDSHA of the uncontrolled system.}
Consider a subcomponent $S$, having the form $S = \sum_{i=1}^k
\ssev{a}_i\pc\alpha_i.S + \ev{init}\pc\alpha.S$. $S$ is a
self-looping agent which can react to events $\ssev{a}_i$ modifying
the state of the influence $\iname{i}_S$, which is specific to $S$,
see Remark~\ref{rem:wellDefined}.

First of all, we need to collect all influences and events appearing
in $S$. The set of influences $\is(S)$ of a subcomponent $S$ is
defined inductively by $\is(\ssev{a}\pc\alpha.S) = \{\alpha\}$ and
$\is(S_1 + S_2) = \is(S_1) \cup \is(S_2)$, while the set of events
$\es(S)$ of $S$ is defined by $\es(\ssev{a}\pc\alpha.S) =
  \{\ssev{a}\}$ if $\ssev{a}\neq \ev{init}$, $\es(\ssev{a}\pc\alpha.S) =
  \emptyset$ otherwise, and $\es(S_1 + S_2) = \es(S_1) \cup
\es(S_2)$.
The set $\is(S)$ contains all the possible flows that can be
generated by the influence with name $\iname{i}_S$. As only one of
them can be active in each state of the system, we will introduce
one mode for each element of $\is(S)$ in the TDSHA of $S$. Moreover,
in each such mode, the only continuous transition will be the one
that can be derived from the corresponding influence. As for
discrete edges, observing that the flat structure of $S$ is such
that the response to all events is always enabled, we will have an
outgoing transition for each event appearing in $S$ in each mode of
the associated TDSHA. The target state of the transition will be the
mode corresponding to the influence following the event. Resets and
guards will be set to $true$, as event conditions will be associated
with the controller. Rates of transitions derived from stochastic
events $\sev{a}\in \Ev_s$  will be set to $\act(\sev{a})$, as
required by the consistency condition of TDSHA. Finally, weights
will be set to 1, while the initial mode will be deduced from the
$\ev{init}$ event.

\begin{figure}[!t]
  \begin{center}
   \includegraphics[height=5cm]{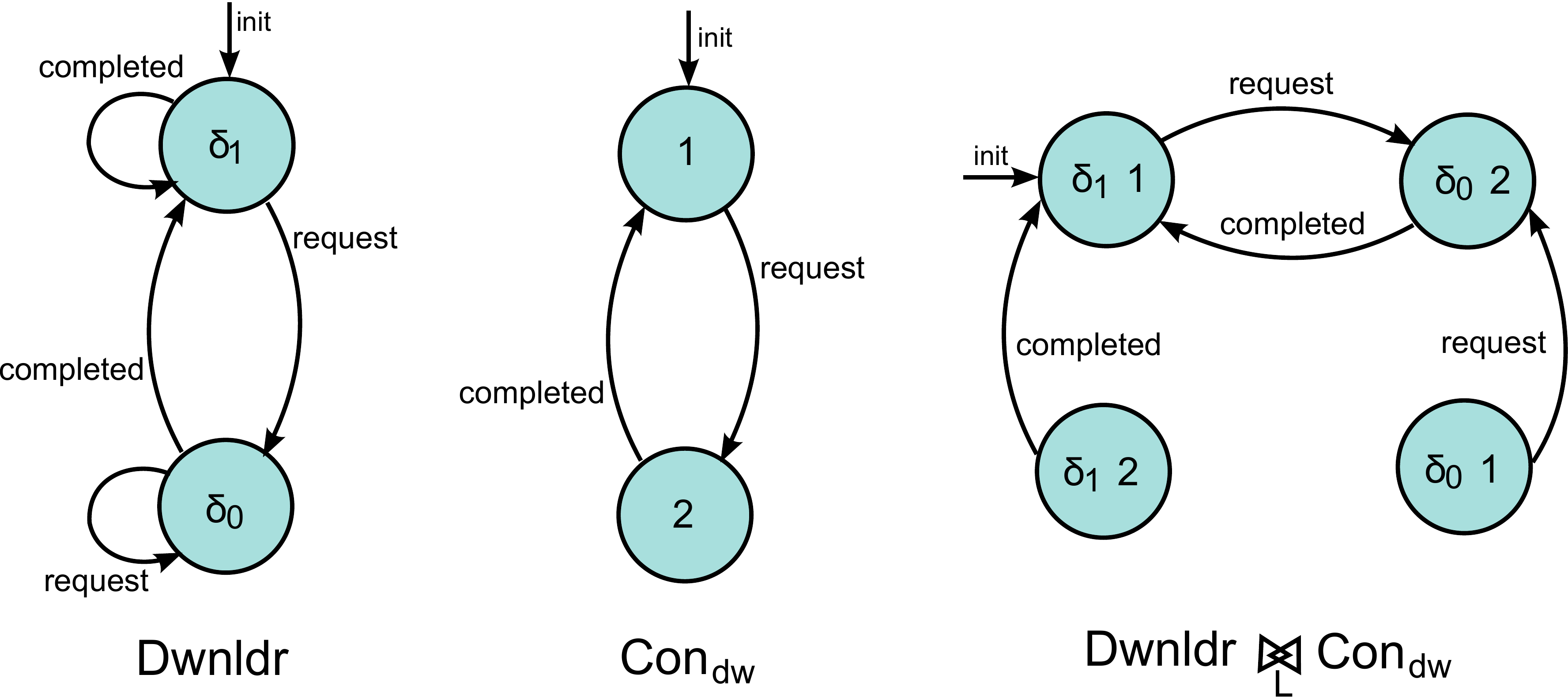}\\
  \end{center}
  \caption{Schematic representation of the TDSHA  $\TDSHA{}(\Dwnldr)$, 
   associated with the download module of the Orbiter, (\textbf{left})
   of the TDSHA $ \TDSHA{}(\Con_\dw)$ associated with the 
   download controller (\textbf{middle}) and of their  TDSHA product
   $\TDSHA{}(\Dwnldr)\otimes_{L}\TDSHA{}(\Con_\dw)$ 
   (\textbf{right}).}\label{fig:ex}
    \vspace{-0.5cm}
\end{figure}

\noindent Consider the subcomponent
\[\Dwnldr \rmdef \ev{init}\pc\delta_1.\Dwnldr
+ \sev{request}\pc\delta_0.\Dwnldr +
\sev{completed}\pc\delta_1.\Dwnldr \]
describing the downloading module of the Orbiter system of
Section~\ref{sec:HYPE}. The TDSHA associated with it is visually
depicted in Figure~\ref{fig:ex} (left). It has two modes,
corresponding to the two different influences $\delta_1 =
(\dw,r,\const)$ and $\delta_0 = (\dw,0,\const)$, and two edges, labeled
by $\sev{request},\sev{completed}$. The initial state is the mode
corresponding to $\delta_1$.

We collect now such considerations into a formal definition.
\begin{definition}\label{def:TDSHAsubcompoenent}
Let $S = \sum_{i=1}^k \ssev{a}_i\pc\alpha_i.S +
\ev{init}\pc\alpha.S$ be a subcomponent of the HYPE model
$(\ConSys,\V,\IN,\IT,$ $\Ev_d,\Ev_s,\Ac,\ec,\iv,\EC,\ID)$. The
TDSHA\\ $\TDSHA{}(S) = \TDSHAtuple{}$ associated with $S$ is defined
by
\begin{enumerate}
  \item $\modes{} = \{\mode{\alpha}~|~\alpha\in\is(S)\}$; $\vrX{} =
  \V$; $\events{} = \Ev_d\cup \Ev_s$;
  \item $\initial{} = (\mode{\alpha},true)$, where $S =
  \ev{init}\pc\alpha.S + S'$;
  \item $\TC{} = \{(\mode{\alpha},\vr{1}_{\iv(\iname{i}_S)},r\cdot\eval{I})~|~\alpha=(\iname{i}_S,r,I)\in\is(S)
  \}$, where $\vr{1}_{\iv(\iname{i}_S)}$ is the vector equal to
  1 for the component corresponding to variable
  $\iv(\iname{i}_S)$ and zero elsewhere;
  \item $\TD{} =
\{(\mode{\alpha_1},\mode{\alpha_2},1,true,true,\ev{a})~|~\ev{a}\in\es(S)\cap\Ev_d\wedge\alpha_1\in\is(S)\wedge
S=\ev{a}\pc\alpha_2.S + S'\}$
  \item $\TS{} =
\{(\mode{\alpha_1},\mode{\alpha_2},true,true,\act(\sev{a}),\sev{a})~|~\sev{a}\in\es(S)\cap\Ev_s\wedge\alpha_1\in\is(S)\wedge
S=\ev{a}\pc\alpha_2.S + S'\}$
\end{enumerate}
\end{definition}

Once we have the TDSHA of all subcomponents, we can build the TDSHA
of the full uncontrolled system by applying the product construction
of TDSHA. We capture this in the following definition.

\begin{definition}\label{def:TDSHAuncontrolled}
$\phantom{a}$
\begin{enumerate}
  \item Let $P = P_1\sync{L}P_2$ be a component. Its TDSHA is defined
  recursively by $\TDSHA{}(P_1\sync{L} P_2) = \TDSHA{}(P_1)\tprod{L}\TDSHA{}(P_2)$.
  \item Let $\Sigma=\Sigma_1\sync{L}\Sigma_2$ be an uncontrolled system. Its TDSHA is defined
  recursively by $\TDSHA{}(\Sigma_1\sync{L}\Sigma_2) =
  \TDSHA{}(\Sigma_1)\tprod{L}\TDSHA{}(\Sigma_2)$.
\end{enumerate}

\end{definition}


\paragraph{TDSHA of the controller.}

Dealing with the controller is simpler, as controllers are
essentially finite state automata which impose causality on the
happening of events. As anticipated at the beginning of the section,
event conditions will be assigned to edges of TDSHA associated with
controllers. Controllers are defined by the two level syntax $M =
\ev{a}.M\mid M+M$ and $Con = M\mid Con\sync{l}Con$, hence sequential
controllers are composed in parallel and synchronized on sets of
actions. As for the uncontrolled system, we will first define the
TDSHA of sequential controllers, and then combine them with the
TDSHA product construction. Note that all events will be properly
dealt with through this construction, as they all appear in the
controller, see Remark~\ref{rem:wellDefined}.

Consider a sequential controller $M = \sum_{i} \ssev{a}_i.M_i$. The
derivative set of $M$ is defined recursively by $ds(M) = \{M\}\cup
\bigcup_i ds(M_i)$, where two summations coincide if they are equal
up to permutation of addends.

\begin{definition}\label{def:TDSHAseqContr} Let
$(\ConSys,\V,\IN,\IT,\Ev_d,\Ev_s,\Ac,\ec,\iv,\EC,\ID)$ be a HYPE
model with sequential controller $M$.
Then $\TDSHA{}(M) = \TDSHAtuple{}$, 
the TDSHA associated with $M$, is defined by
\begin{enumerate}
  \item $\modes{} = \{\mode{M'}~|~M'\in ds(M)\}$; $\vrX{} = \V$; $\events{} = \Ev_d\cup \Ev_s$;
  \item $\initial{} = (\mode{M},\rs(\ev{init}))$, where
  $\rs(\ev{init})$ is the reset associated with the $\ev{init}$
  event.
  \item $\TC{} = \emptyset$;
  \item $\TD{} = \{(\mode{M_1},\mode{M_2},1,\act(\ev{a}),\rs(\ev{a}),\ev{a})
  ~|~M_1 = \ev{a}.M_2,\ M_1,M_2\in ds(M),\ \ev{a}\in\Ev_d,\
  \ec(\ev{a}) = (\act(\ev{a}),$ $\rs(\ev{a})) \}$;
  \item  $\TS{} = \{(\mode{M_1},\mode{M_2},true,\rs(\sev{a}),\act(\sev{a}),\sev{a})
  ~|~M_1 = \sev{a}.M_2,\ M_1,M_2\in ds(M),\ \sev{a}\in\Ev_s,\
  \ec(\sev{a}) = (\act(\sev{a}),$ $\rs(\sev{a})) \}$, where
  $\act(\sev{a}):\bbR^{|\V|}\rightarrow\bbR^+$ is the rate of
  the transition;
\end{enumerate}

\end{definition}

\begin{definition}\label{def:TDSHAcontroller}
Let $Con = Con_1\sync{L}\Con_2$ be a controller. The TDSHA of $Con$
is defined recursively as $\TDSHA{}(Con) =
\TDSHA{}(Con_1)\tprod{L}\TDSHA{}(Con_2)$.
\end{definition}

The product construction of Definitions~\ref{def:TDSHAuncontrolled}
and~\ref{def:TDSHAcontroller} can be carried on because the factors
TDSHA are reset-compatible and init-compatible. This is trivial both
for the uncontrolled system (all resets are $true$) and for the
controller (resets for the same event are equal). Furthermore,
stochastic transitions have consistent rates, as their rate depends
only on the labelling event.

Consider the controller of the download module of the orbiter; its
TDSHA is depicted in Figure~\ref{fig:ex} (middle), omitting the
explicit representation of rates and resets.

\paragraph{TDSHA of the HYPE model.}

Once we have built the TDSHA of the controller and of the
uncontrolled system, we simply have to take their product.

\begin{definition}\label{def:TDSHAofHYPE}
Let $(\ConSys,\V,\IN,\IT,\Ev_c,\Ev_s,\Ac,\ec,\iv,\EC,\ID)$ be a
HYPE model, with controlled system
$\ConSys = \Sigma\sync{L} \ev{init}.\Con$. The
TDSHA associated with $\calM$ is
$$\TDSHA{}(\calM) = \TDSHA{}(\Sigma)\tprod{L}\TDSHA{}(\Con).$$
\end{definition}

\begin{figure}[!t]
  \begin{center}
   \includegraphics[height=4.5cm]{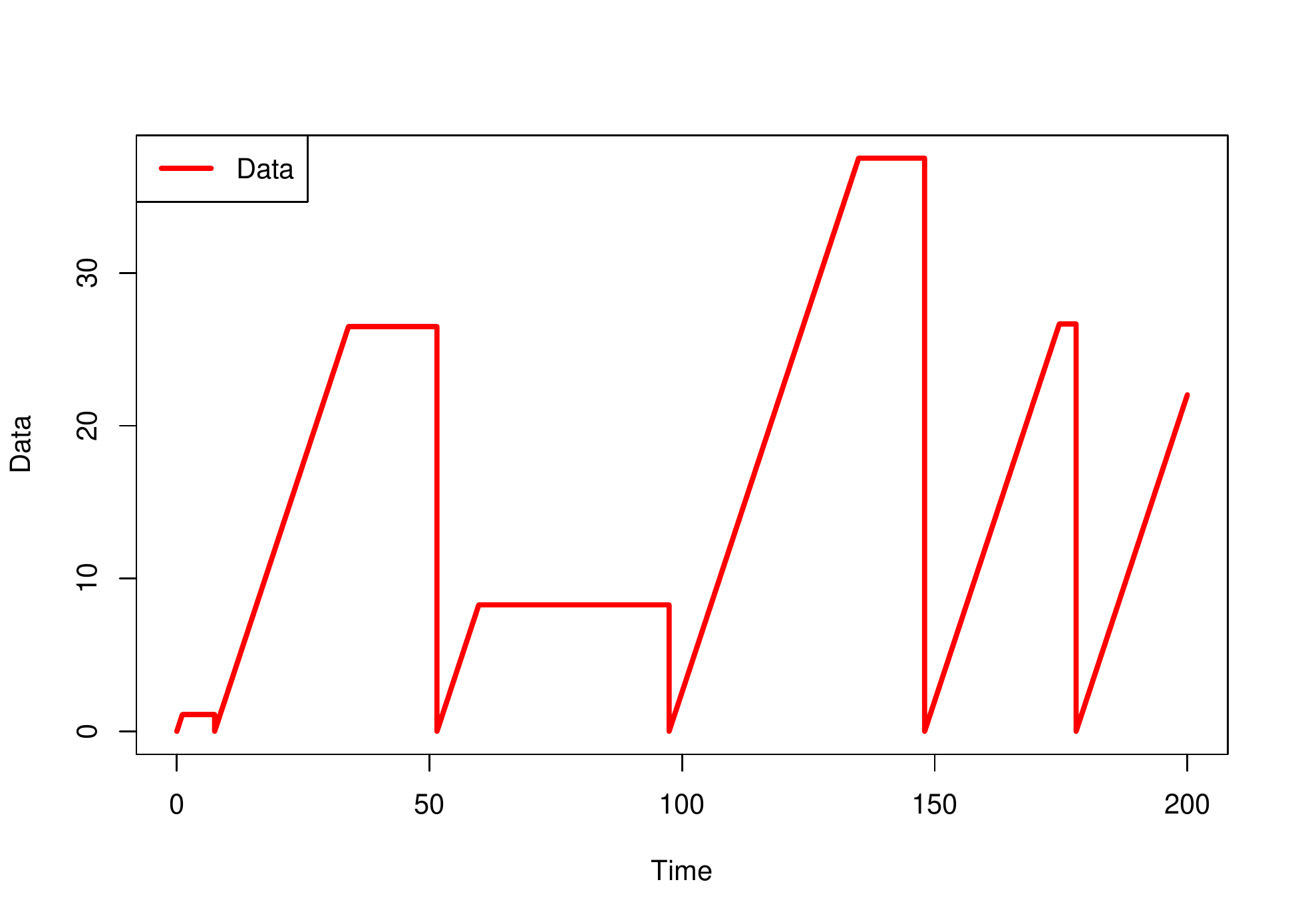}
    \vspace{-0.8cm}
  \end{center}
  \caption{Sampled trajectory of the accumulated data of the extended
  orbiter model of Section~\ref{sec:stochHYPE}. Data increases during accumulation phases, and remains constant
  during downloads. It is erased right after the download finished.
  Rate values, fixed just for illustrative purposes, are $r = 1.0$,
  $\lambda_r = 0.04$, $\lambda = 0.5$, $\mu = 10.0$.}\label{fig:orbDyn}
  \vspace{-0.5cm}
\end{figure}

\begin{example}\label{ex:4}
In Figure~\ref{fig:ex} (right) we show the product
$\TDSHA{}(Dwnldr)\otimes_{L}\TDSHA{}(\Con_\dw)$,
$L=\{\sev{request},\sev{completed}\}$, in order to give an idea of
the product construction. In Figure~\ref{fig:orbDyn}, instead, we
show a trajectory of the variable $D$, describing the amount of data
collected. As we can see, periods in which the data is collected
(linearly), are interleaved by downloads, in which data is not
accumulated. Once the download has finished, $D$ is set back to
zero. Both the download time and the periods between two consecutive
downloads are randomly distributed.
\end{example}

As already evident from the previous example, the construction we
have defined actually generates TDSHA  with many \emph{unreachable
states}. This is a consequence of the fact that sequentiality and
causality on actions is imposed just on the final step, when the
controller is synchronized with the uncontrolled system. Once the
TDSHA is constructed, however, it can be pruned by removing
unreachable states (the TDSHA of Figure~\ref{fig:ex} (right) has
indeed just two reachable states from the initial one). In order to
limit combinatorial explosion, one can prune TDSHA's at each
intermediate stage. A formal definition of this policy, however,
would have made the mapping from HYPE to TDSHA much more complex.

\begin{figure}[!h]
  \begin{center}
   \includegraphics[height=4.5cm,width=6cm]{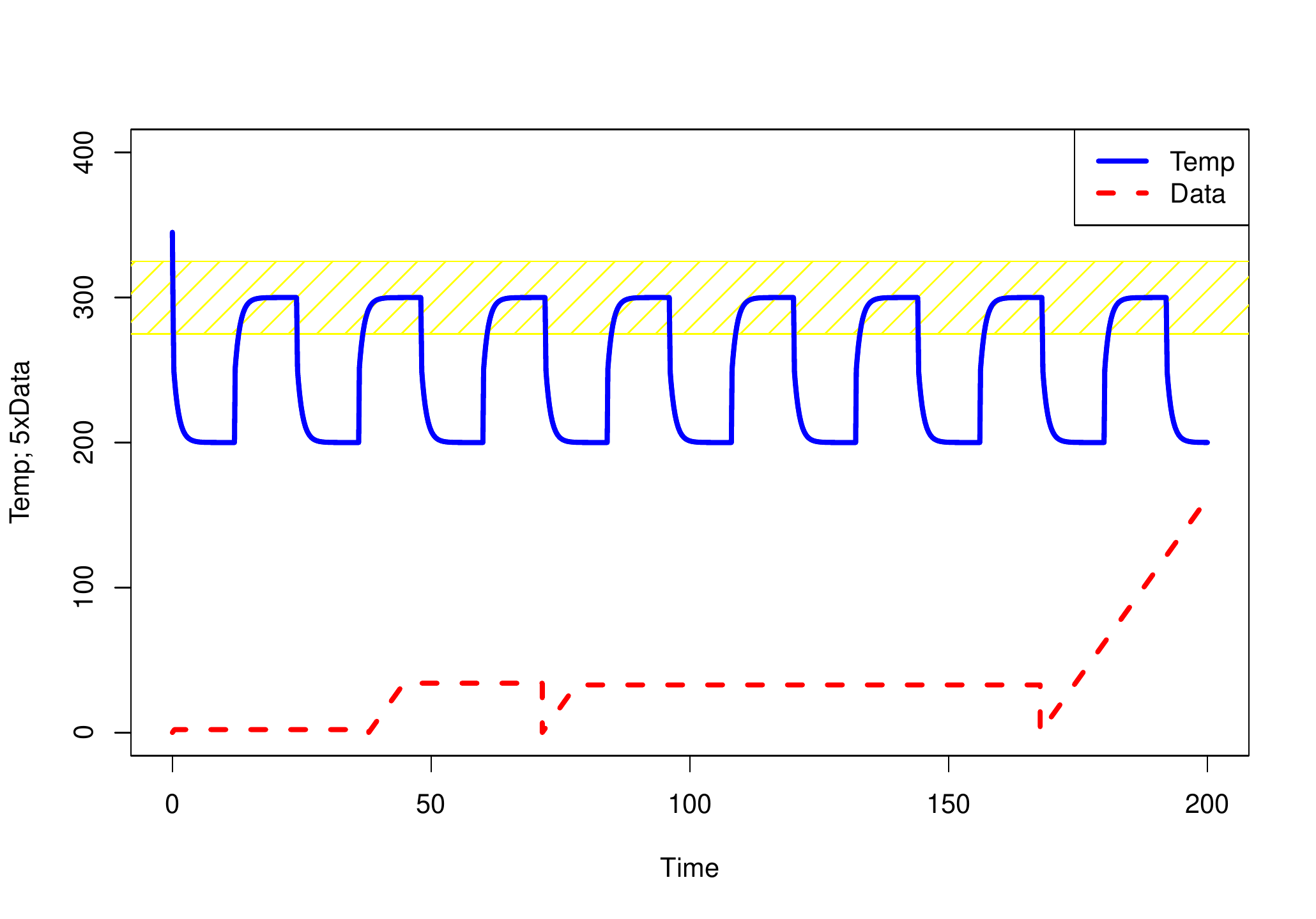}
   \includegraphics[height=4.5cm,width=6cm]{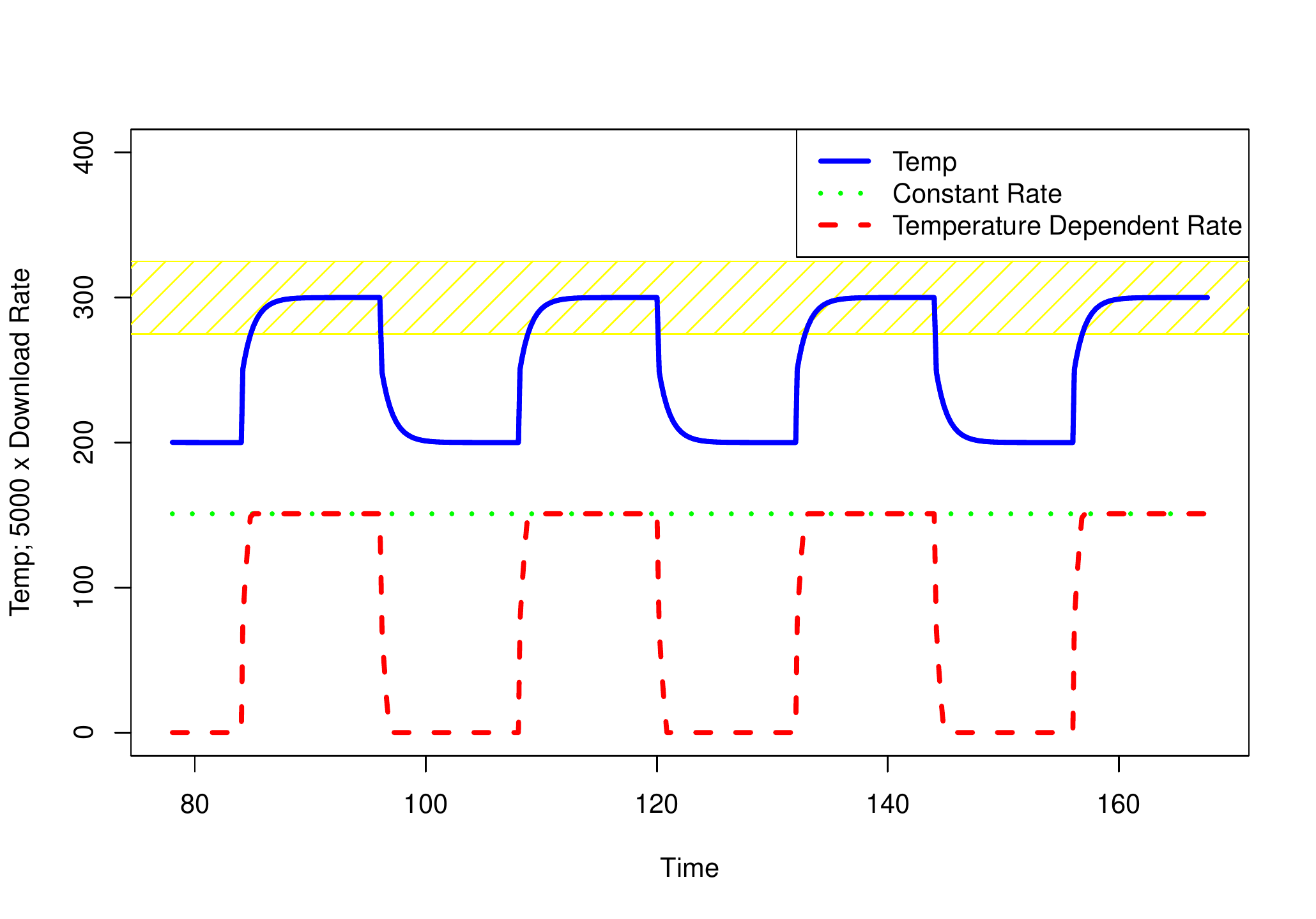}\vspace{-0.5cm}\\
  \end{center}
  \caption{ (\textbf{left}) Sampled trajectory of the accumulated data of the extended
  orbiter model of Section~\ref{sec:stochHYPE}, with download rate depending on temperature.
  The rate is maximal, and equal to $\frac{\lambda}{\mu + D}$ when temperature is in the operational regime,
  in this example, when $275\leq K\leq 325$. When the temperature is lower than 275 (higher than 325), the download rate linearly decreases
  to 0, reaching it when $K= 225$ ($K=350$). Rate values, for the downloader are $r = 1.0$,
  $\lambda_r = 0.1$, $\lambda = 1.0$, $\mu = 10.0$. Rates  and parameters for the temperature control mechanism
  are $f_h = 200$, $r_d = 100$, $r_s = 400$, $k_1 = k_2 = 250$, $k_3=k_4 =300$.
  The download time becomes longer with respect to Figure~\ref{fig:orbDyn}, as the temperature falls repeatedly
  below the operational regime.
  (\textbf{right}) Plot of the downloading rate when it is constant (green) or when it depends on temperature (red).
  In the latter case, the rate is periodically reduced to zero, as temperature falls below 225. The shadowed region indicates, in both plots,
  the interval of temperatures in which the download has maximum speed.}\label{fig:triggerTime}
  \vspace{-0.3cm}
\end{figure}

\paragraph{Orbiter revisited.} We consider now a more complex
version of the orbiter, in which the download time depends also on
the current temperature. The operational speed of the download can
be reduced linearly down to zero if the temperature is too high or
too low. In order to implement such a modification, we simply have
to modify the rate function in the event condition of event
$\sev{completed}$, replacing it with a suitable function of
accumulated data and temperature. A sampled trajectory is shown in
Figure~\ref{fig:triggerTime}~(left), while in
Figure~\ref{fig:triggerTime}~(right) we show how the firing time of
$\sev{completed}$ depends on temperature.

%% file: related.tex
\section{Related Work}\label{seC:relatedWork}

The modelling approach of HYPE, based on the composition of
individual flows, makes it different from other hybrid process
algebras~\cite{KhadK:06a} and from hybrid automata~\cite{HenzH:96a}.
In these other approaches the continuous dynamics is
specified by embedding ODEs within the syntactic description of
models, while in HYPE, ODEs emerge as a combination of active
flows. A more detailed comparison between HYPE and other hybrid
modelling formalisms can be found in~\cite{CONCUR09,HYPE-journal}.

As far as stochastic hybrid systems are concerned, there has  been
previous work aimed at  making modelling compositional.  In
\cite{CPDP:Stru-Jul-Scha:2003}, Strubbe \emph{et al.} introduce
Communicating Piecewise Deterministic Markov Processes (CPDP).  This
is an automata based formalism which models a system as interacting
automata.  Their chosen level of abstraction is somewhat lower level
than ours, comparable with TDSHA. In CPDP, as in HYPE, instantaneous
transitions may be triggered either by conditions of the continuous
variables (boundary-hit transitions) or by the expiration of a
stochastic determined delay (Markov transitions).  Interaction
between automata is based on \emph{one-way} synchronisation: in each
interaction one partner is active while the other is passive. In
HYPE, instead, all components may be regarded as active with respect
to each transition in which they participate, as activation
conditions are specified uniquely in the model. Components
participating in a discrete transition are determined by the
construction of the HYPE model, where the synchronisation set $L$ in
$\sync{L}$ specifies which actions must be shared.

The synchronization mechanics of CPDP has been extended in
 \cite{CPDP-ext:Stru-Scha:2005}, introducing an  operator
which  exploits all possible interactions of active and passive
actions. In \cite{CPDP:bisim:2005} the authors define a notion of
bisimulation for both PDMPs and CPDPs and show that if CPDPs are
bisimilar then they give rise to bisimilar PDMPs.  Furthermore the
equivalence relation is a congruence with respect to the composition
operator of CPDPs.

%% file: conc.tex
\section{Conclusions}\label{sec:conc}

In this paper we extended the hybrid process algebra HYPE, allowing
events to fire at (exponentially distributed) random times. Although
from a syntactic point of view the modifications with respect to the
original version of HYPE are minimal (non-urgent events become
stochastic by replacing their activation condition  $\bot$  with a
functional rate), the semantics of the language is considerably
enriched. The stochastic hybrid systems obtained from HYPE models
fall in the class of Piecewise Deterministic Markov Processes. In
the paper, we concentrated on showing how such a semantics can be
defined. We used an intermediate formalism, namely Transition-Driven
Stochastic Hybrid Automata, which can then be mapped to PDMPs. The
way we defined the semantics in terms of TDSHA is quite different
from the original definition of~\cite{CONCUR09,HYPE-journal}, in
which a hybrid automaton is extracted from the labeled transition
system of a HYPE model, defined according to a suitable operational
semantics. Here, instead, we directly manipulate the model at the
syntactic level.

The mapping from TDSHA to PDMP is quite straightforward, except in
one point: One has to check that the HYPE model is \emph{well-behaved},
meaning that it is not possible that an infinite sequence of
instantaneous transitions fires in the same time instant.  Unfortunately,
checking this property in general is undecidable, hence
in~\cite{HYPE-journal} we put forward a set of decidable but stricter
conditions on HYPE models, that guarantee that a model is well-behaved
and that are usually satisfied in practical cases.

As the syntax of HYPE is basically unchanged, all the results
of~\cite{CONCUR09,HYPE-journal} depending on syntactic features
still hold. In particular, the notion of bisimulation of HYPE models
extends untouched in this new setting. As a future investigation, we
plan to compare this bisimulation relation with other bisimulations
designed for PDMPs~\cite{CPDP:bisim:2005}.

In the current version of HYPE, stochasticity has been introduced
just in terms of random occurrence in the time of events. It is
often useful to have stochasticity also in resets. This would allow
the quantitative modelling of uncertainty in the outcome of certain
actions. Such an extension can be done along the lines of the
current paper, even if it requires a modification of the definition
of TDSHA, allowing stochastic resets. However, the class of target
stochastic processes remains that of PDMP.

Future work includes also the implementation of an efficient
simulator for (stochastic) HYPE. Moreover, we will model specific
case studies, to prove its effectiveness as a hybrid modelling
language.

\paragraph{Acknowledgements\ \ } This work was supported by Royal Society
International Joint Project (JP090562).  Vashti Galpin is supported
by the EPSRC SIGNAL Project, Grant EP/E031439/1. Luca Bortolussi is
supported by GNCS. Jane Hillston has been supported by EPSRC under ARF
EP/c543696/01.